\documentclass[12pt]{article}
\pdfoutput = 1
\usepackage{amsmath,amssymb,bm,epsfig,afterpage}
\usepackage{mathtools}
\usepackage{cite}
\usepackage{here}
\usepackage{color}
\usepackage{tabularx}
\usepackage{simplewick} 
\usepackage{bbm}
\usepackage[samesize]{cancel} 
\usepackage{comment}
\usepackage{nicematrix}

\renewcommand{\thefootnote}{\fnsymbol{footnote}}
\newcommand{\vev}[1]{{\langle{#1}\rangle}}

\newcommand{\br}[2]{\text{Br}({#1}\to{#2})}

\newcommand{\abs}[1]{\left|{#1}\right|}

\newcommand{\eps}{\epsilon}

\newcommand{\order}[1]{\mathcal{O}\left({#1}\right)}

\newcommand{\pr}{\prime}

\newcommand{\Lcal}{\mathcal{L}}
\newcommand{\Mcal}{\mathcal{M}}

\newcommand{\GeV}{\mathrm{GeV}}
\newcommand{\TeV}{\mathrm{TeV}}

\newcommand{\diagp}[1]{\mathrm{diag}\left({#1}\right)}

\newcommand{\ygy}[3]{y_{#1}^\dag g_{#2} y_{#3}}
\newcommand{\yy}[2]{y_{#1}^\dag y_{#2}}

\newcommand{\la}{{\lambda}}
\newcommand{\ka}{{\kappa}}

\newcommand{\ol}[1]{\overline{#1}}

\newcommand{\SM}{\mathrm{SM}}

\newcommand{\ds}[1]{\cancel{#1}}

\definecolor{darkviolet}{rgb}{0.58, 0.0, 0.83}

\newcommand{\dsq}{\ds{q}\,}

\newcommand{\dsdel}{\ds{\partial}\,}

\newcommand{\rel}{\mathrm{rel}}
\newcommand{\vrel}{v_{\mathrm{rel}}} 
\newcommand{\eq}{\mathrm{eq}}

\newcommand{\fo}{\mathrm{fo}}

\numberwithin{equation}{section}

\allowdisplaybreaks[3]
\newcolumntype{Y}{&gt;{\centering\arraybackslash}X} 

\usepackage{hyperref}

\definecolor{asparagus}{rgb}{0.53, 0.66, 0.42}
\definecolor{darkspringgreen}{rgb}{0.09, 0.45, 0.27}
\definecolor{darkturquoise}{rgb}{0.0, 0.81, 0.82}
\definecolor{dollarbill}{rgb}{0.52, 0.73, 0.4}

\begin{document}

\begin{titlepage}

\begin{flushright}
 {\tt
}
\end{flushright}

\vspace{1.2cm}
\begin{center}
{\Large
{\bf
  Taming lepton portal dark matter 
  by a non-invertible selection rule 
}
}
\vskip 2cm
Junichiro Kawamura$^a$~\footnote{kawamura-ju@kumamoto-hsu.ac.jp} 
\vskip 0.5cm

{\it $^a$
  Faculty of Health Science,
  Kumamoto Health Science University,  \\ 
  Kumamoto, 861-5598, Japan 
}\\[3pt]

\vskip 1.5cm

\begin{abstract}
We point out that a non-invertible selection rule can control the flavor
violation inherent in lepton portal dark matter, a weakly interacting massive
particle candidate that can evade severe constraints from direct detection
experiments.
We show that ordinary invertible symmetries cannot suppress
flavor-violating portal couplings while accommodating the large mixing angles
in the PMNS matrix.
We then present a simple construction based on the $\mathbb{Z}_2$ gauging
of a $\mathbb{Z}_5$ symmetry, which realizes one-flavor dominance of the portal
couplings consistently with the PMNS mixing matrix.
We also discuss possible flavor leakage in a type-I seesaw completion
and study experimental probes of the model through lepton observables,
especially lepton flavor non-universality in $Z$ boson decays.
\end{abstract}
\end{center}
\end{titlepage}

\clearpage

\renewcommand{\thefootnote}{\arabic{footnote}}
\setcounter{footnote}{0}

\tableofcontents
\clearpage 

\section{Introduction}

Dark matter (DM) is one of the most important problems in modern particle physics and cosmology.
Among many possibilities, weakly interacting massive particles (WIMPs) are plausible candidates,
since their relic abundance $\Omega h^2 \simeq 0.12$~\cite{Planck:2018vyg} can be explained
by thermal freeze-out~\cite{Jungman:1995df,Bertone:2004pz,Arcadi:2017kky,Roszkowski:2017nbc}.
WIMPs have been extensively explored by direct detection experiments
looking for their scattering with nucleons or electrons.
The null results from these experiments have placed severe constraints
on the DM scattering cross sections,
especially for DM-nucleon scattering above the GeV scale~\cite{XENON:2023cxc,LZ:2022lsv,PandaX:2024qfu}.
As a result, a wide class of WIMP models has been strongly constrained under standard assumptions.

Lepton portal dark matter (LPDM)~\cite{Bai:2014osa,Kawamura:2020qxo} is a well-motivated WIMP scenario
in which the direct detection rate can be naturally suppressed.
In this class of models, DM couples to the SM leptons
through Yukawa interactions with vector-like leptons.
The DM annihilation proceeds through the $t$-channel exchange of the vector-like leptons,
and the observed relic abundance can be reproduced for portal Yukawa couplings
of $\order{1}$.
Since the DM does not couple directly to quarks,
the scattering with nucleons occurs only at the 2-loop level for real scalar DM,
and is therefore far below current direct detection sensitivities~\cite{Garani:2021zrr}.

Despite this attractive feature,
LPDM generically suffers from a flavor problem.
The portal Yukawa couplings can involve all three flavors of the SM leptons
and hence induce lepton flavor violation, such as $\mu\to e\gamma$.
In the literature, it is often simply assumed that DM couples
to only one lepton flavor in order to avoid this problem.
However, as we show below, such one-flavor dominance cannot be justified
by ordinary invertible symmetries, such as Abelian discrete symmetries $\mathbb{Z}_N$,
if the large mixing angles in the PMNS matrix are realized
without introducing additional spurion fields.

The aim of this paper is to show that a non-invertible selection rule
can tame the flavor violation of LPDM without conflicting with the PMNS mixing matrix.
Non-invertible symmetries have recently been developed
as a generalization of ordinary symmetries in quantum field theory~\cite{Schafer-Nameki:2023jdn,Choi:2022jqy}.
As in the case of ordinary symmetries,
they can impose selection rules on interactions in the Lagrangian. 
However, the resulting selection rules are not described by group charges with ordinary inverse elements.
In this work, we use the non-invertible selection rule
obtained by gauging a $\mathbb{Z}_2$ automorphism of a $\mathbb{Z}_N$ symmetry.
We show that a simple realization based on $\mathbb{Z}_5$
achieves one-flavor dominance of the LPDM portal couplings 
while keeping a viable lepton flavor structure.
Such selection rules may arise in low-energy effective theories of extra-dimensional setups~\cite{Kobayashi:2024yqq,Funakoshi:2024uvy,Dong:2025pah,Kobayashi:2025znw,Kobayashi:2025ocp}.

The rest of this paper is organized as follows.
In Section~\ref{sec-LPDM},
we introduce the LPDM model and discuss its thermal relic density,
existing DM and collider constraints, and the flavor problem of the portal couplings.
In Section~\ref{sec-NISR}, we show that ordinary invertible symmetries
cannot realize one-flavor dominance of the portal couplings consistently with the PMNS mixing matrix,
and then present a realization based on a non-invertible selection rule.
We also discuss possible flavor leakage in a type-I seesaw completion.
In Section~\ref{sec-pheno}, we study lepton observables,
focusing on anomalous magnetic moments and lepton flavor non-universality in $Z$ boson decays.
Section~\ref{sec-concl} is devoted to the conclusion.
The details of the 1-loop calculations are summarized in
Appendix~\ref{app-formula}.

\section{Lepton portal dark matter}
\label{sec-LPDM}

In this section,
we introduce the lepton portal dark matter (LPDM) model~\cite{Bai:2014osa,Kawamura:2020qxo}.
We discuss the annihilation cross section relevant for thermal freeze-out,
summarize the existing DM and collider constraints,
and explain the flavor problem of the portal couplings.

\subsection{Model}

\begin{table}
  \centering
  \begin{tabular}[t]{c|ccc|ccc} \hline 
               & $X$ & $L$ & $E$ & $\ell_i$  & $e_i$ & $H$  \\ \hline\hline 
    $SU(2)_L$  & 1   & 2   & 1   & 2         & 1     & 2    \\
    $U(1)_Y$   & 0   &$-1/2$ & $-1$  & $-1/2$ & $-1$    & $1/2$ \\  \hline
$\mathbb{Z}_2$ & $-$ & $-$   & $-$   & $+$         & $+$     & $+$   \\  \hline
  \end{tabular}
  \caption{\label{tab-contents}
    Matter contents of the LPDM model.
    The flavor index of the SM leptons $\ell_i$ and $e_i$ runs over $i=1,2,3$. 
  }
\end{table}

The matter contents of the model are summarized in Table~\ref{tab-contents}.
In addition to the Standard Model (SM) particles,
there are the real scalar DM $X$ 
and vector-like leptons $L$ and $E$
whose gauge charges are the same as the SM leptons $\ell_i$ and $e_i$, respectively.  
The $SU(2)_L$ doublets are defined as
\begin{align}
  \label{eq:doublets}
  H =
  \begin{pmatrix}
    H^+ \\ H^0
  \end{pmatrix},
  \quad
  \ell_i = 
  \begin{pmatrix}
    \ell^0_i \\ \ell^-_i 
  \end{pmatrix},
  \quad
  L = 
  \begin{pmatrix}
    L^0 \\ L^- 
  \end{pmatrix},   
\end{align}
where the superscripts denote their electric charges.
We impose a $\mathbb{Z}_2$ symmetry
under which the new particles are odd,
ensuring the stability of the lightest $\mathbb{Z}_2$-odd particle,
which we take to be the DM $X$.

The relevant mass terms and Yukawa interactions are given by 
\begin{align}
  -\Lcal =&\ \frac{1}{2} m_X^2 X^2
           + m_L \ol{L} L + m_E \ol{E}E
            \\ \notag 
          &\ + \left(
             -  y_e^{ij} \ol{e}_i P_L \ell_j \cdot \tilde{H} 
             + \la^i_L X \ol{L} P_L \ell_i
             + \la^i_E X \ol{e}_i P_L E 
             - \kappa \ol{E} P_L L\cdot \tilde{H} + \ol{\kappa} \ol{L} P_L E H + h.c. \right),  
\end{align}
with $\tilde{H} := i\sigma_2 H^* = (H^{0*}, -H^-)$.
For example, the products of the doublets are given by  
\begin{align}
  \label{eq:prods}
  &  \ol{L} \ell = \ol{L}^0 \ell^0 + \ol{L}^+ \ell^-,
    \quad 
    L\cdot \tilde{H} = - L^0 H^- - L^- H^{0*},
\end{align}
where $\cdot$ represents the contraction with $i\sigma_2$ in $SU(2)_L$ space.
The SM Lagrangian terms responsible for the lepton masses are given by 
\begin{align}
 - \Lcal_{\SM} = &\
           -   y_e^{ij} \ol{e}_i P_L \ell_j \cdot \tilde{H}                    
            + \frac{c^{ij}_n}{2\Lambda_n}
            (\ol{\ell}^{c}_i \cdot H) P_L (\ell_j \cdot H) + h.c.,     
\end{align}
where $c_n^{ij} = c_n^{ji}$ are dimensionless constants and $\Lambda_n$ is the cut-off scale.
Here, we consider the dimension-5 effective Weinberg operator for the neutrino masses.

After the electroweak symmetry breaking by $\vev{H^0}=v_H$,
the SM lepton masses are given by
\begin{align}
  \left[ \Mcal_e \right]_{ij} := y^{ij}_e  v_H,
  \quad
  \left[ \Mcal_n \right]_{ij} := c^{ij}_n \frac{v_H^2}{\Lambda_n}.    
\end{align}
The charged vector-like lepton mass matrix for $(L^-,E)$ is given by
\begin{align}
  \Mcal_E =
  \begin{pmatrix}
    m_L & \ol{\kappa} v_H \\ \kappa v_H & m_E 
  \end{pmatrix},
\end{align}
and the neutral component $L^0$ has mass $m_L$.
The mass matrices are diagonalized by unitary matrices as 
\begin{align}
  V_R^\dagger \Mcal_e V_L = \mathrm{diag} \left(m_e, m_\mu, m_\tau \right),
  \quad
  U^T_n \Mcal_n U_n = \mathrm{diag}\left(m_{n_1}, m_{n_2}, m_{n_3}\right),
\end{align}
and 
\begin{align}
 U_R^\dagger \Mcal_E U_L = \mathrm{diag}\left(m_{E_1}, m_{E_2}\right). 
\end{align}
The Dirac fermions in the mass basis are given by
\begin{align}
  \psi_{e} =  V_L^\dag P_L \ell^- + V_R^\dag P_R e,
  \quad
  \psi_{n} =  U_n^\dag P_L \ell^0, 
  \quad
  \begin{pmatrix}
    \psi_{E_1} \\ \psi_{E_2} 
  \end{pmatrix}
  =  \left( U_L^\dag P_L + U_R^\dag P_R \right)
  \begin{pmatrix}
    L^- \\ E 
  \end{pmatrix},
\end{align}
where the SM flavor indices are omitted. 
We parametrize the unitary matrices of the vector-like leptons as
\begin{align}
  \label{eq:ULUR}
  U_L =
  \begin{pmatrix}
    c_L & s_L \\ -s_L^* & c_L 
  \end{pmatrix},
  \quad
  U_R =
  \begin{pmatrix}
    c_R & s_R \\ -s_R^* & c_R 
  \end{pmatrix},  
\end{align}
with $c_X^2 + \abs{s_X}^2 = 1$ for $X=L,R$.   
The PMNS matrix is then given by
\begin{align}
  U_{\mathrm{PMNS}} = V_L^\dagger U_n.
\end{align}

The portal Yukawa couplings in the mass basis are given by
\begin{align}
  \label{eq:Lymass}
  -\Lcal_{X} = X \sum_{a=1}^2\sum_{i=1}^3
      \ol{\psi}_{E_a}  \left( y_L^{ai} P_L + y_R^{ai} P_R  \right)\psi_{e_i} + h.c., 
\end{align}
where
\begin{align}
  \label{eq-yLyR}
  y_L^{ai} := \left( U_R^\dag \right)_{a1} \sum_{k=1}^3 \la^k_L \left( V_L \right)_{ki},
  \quad
  y_R^{ai} := \left( U_L^\dag \right)_{a2} \sum_{k=1}^3 \la^{k*}_E \left( V_R \right)_{ki}.
\end{align}
The $Z$ boson gauge couplings of the vector-like leptons are given by  
\begin{align}
  \Lcal_{Z} =  Z_\mu \sum_{X=L,R} \Bigl(
  \sum_{i=1}^3 \ol{\psi}_{e_i} \gamma^\mu g_X^e  P_X \psi_{e_i}
  +
  \sum_{a,b=1}^2 \ol{\psi}_{E_a} \gamma^\mu g_X^{ab}  P_X \psi_{E_b} 
 \Bigr),
\end{align}
where
\begin{align}
  \label{eq:Zc}
  g_X^e :=  \frac{g}{c_W} \left( -\frac{1}{2}\delta_{XL} +s_W^2 \right),
  \quad
  g_X^{ab}   :=  \frac{g}{c_W} \left[ -\frac{1}{2}U_X^\dag H_L U_X +s_W^2 \mathbf{1}_2 \right]_{ab}  
\end{align}
with $H_L := \mathrm{diag}(1,0)$.

We note that the scalar DM can couple to the SM Higgs doublet
through the quartic coupling $X^2 \abs{H}^2$ in the scalar potential.
The model reduces to the Higgs portal
DM~\cite{Kanemura:2010sh,Djouadi:2011aa,Djouadi:2012zc,Escudero:2016gzx,Ellis:2017ndg,Athron:2018ipf}
if this coupling dominantly controls the DM annihilation.
Throughout this work, we focus on the case
in which the DM annihilation is dominated by the $t$-channel exchange
of the vector-like leptons discussed in the next section.

\subsection{Annihilation cross section and relic density}

\begin{figure}[t] 
  \centering
  \begin{minipage}{0.48\textwidth}
    \centering
  \includegraphics[width=0.90\linewidth]{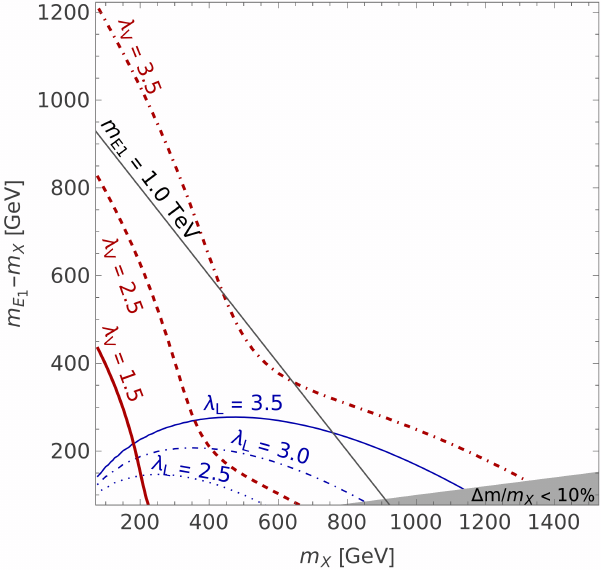}    
  \end{minipage}
  \begin{minipage}{0.48\textwidth}
    \centering    
  \includegraphics[width=0.90\linewidth]{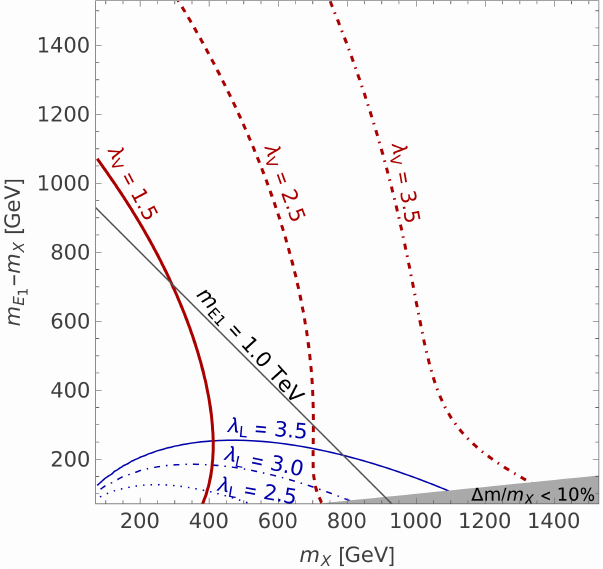}    
  \end{minipage}
  \caption{\label{fig-lambda}
   Portal Yukawa couplings required to reproduce the observed DM relic density
$\Omega_X h^2 = 0.12$.
The left and right panels show the results for
$s_L = s_R = 0.1$ and $0.5$, respectively.
The blue contours show the $d$-wave dominated case with $\lambda_E = 0$,
while the red contours show the case with
$\lambda_V := \lambda_L = \lambda_E$,
where the $s$-wave contribution is present.
We take
$\Delta m_E := m_{E_2} - m_{E_1} = 100~\mathrm{GeV}$.
The gray region indicates
$(m_{E_1} - m_X)/m_X < 10\%$,
where co-annihilation becomes relevant.
  }
\end{figure}

The cross section of the annihilation process $XX \to \ol{\ell}_i \ell_i$
is expanded by the relative velocity of the DM, $v_\rel$, as 
\begin{align}
\sigma \vrel = a + b \vrel^2 + c \vrel^4 + \order{\vrel^6}.  
\end{align}
The three terms correspond to the \(s\)-, \(p\)-, and \(d\)-wave contributions, respectively.
We focus on the case in which the mass difference $m_{E_1} -  m_X$
is so large that the co-annihilation processes do not contribute to the DM annihilation.
The co-annihilation region is intensively studied in Ref.~\cite{Kawamura:2020qxo}.

In the LPDM, the coefficients are given by~\cite{Bai:2014osa,Kawamura:2020qxo}
\begin{align}
  \label{eq:abc}
  a =&\ \frac{1}{\pi m_X^2} \abs{\sum_{a=1,2} y^{ai*}_L y_R^{ai} \frac{\sqrt{r_a}}{1+r_a}  }^2,
  \\ \notag
  b =&\ - \frac{1}{3\pi m_X^2}\Biggl[\left( 
       \sum_{a=1,2} \frac{r_a(1+3r_a)}{(1+r_a)^4} \abs{y^{ai*}_L y^{ai}_R}^2  \right)
\\ \notag        
     &\ \hspace{2.0cm}
       + \mathrm{Re}\left(y^{1i*}_L y^{1i}_R y^{2i}_L y^{2i*}_R  \right)
        \frac{\sqrt{r_1r_2}}{(1+r_1)(1+r_2)}
      \left(
     \sum_{a=1,2} \frac{1+3r_a}{(1+r_a)^2}
      \right)
         \Biggr], 
  \\ \notag
  c =&\ \frac{1}{60\pi m_X^2} \left[
       \frac{\abs{\la_L^{i}}^4} {(1+r_n)^4} +  
       \left( \sum_{a=1,2}\frac{\abs{y^{ai}_L}^2}{(1+r_a)^2} \right)^2
     + \left( \sum_{a=1,2}\frac{\abs{y^{ai}_R}^2}{(1+r_a)^2} \right)^2        
       \right] + \order{ (y_L^{ai}y_R^{bi})^2}, 
\end{align}
where $r_n = m_L^2/m_X^2$ and $r_a = m_{E_a}^2/m_X^2$ for $a=1,2$.
Here, we neglect the SM lepton masses.  
The $s$-wave and $p$-wave contributions appear when $\la_L\la_E \ne 0$,
otherwise the $d$-wave is the leading one.

The number density of the DM, $n_X$, obeys the Boltzmann equation,
\begin{align}
  \label{eq:boltzn}
  \frac{d n_X}{dt} + 3H n_X = - \vev{\sigma v_\rel} (n^2_X - n_{X,\eq}^2),   
\end{align}
where $H$ is the Hubble parameter, 
and $n_{X,\eq}$ is the number density in the thermal equilibrium.
Assuming the Maxwell-Boltzmann distribution and that the DM is non-relativistic,
the thermally averaged cross section is given by
\begin{align}
  \vev{\sigma v_\rel} = a + \frac{6}{x} b + \frac{60}{x^2} c,   
\end{align}
where $x := m_X/T$.
Assuming radiation domination at freeze-out, the DM yield $Y_X:=n_X/s$,
with $s$ being the entropy density, obeys
\begin{align}
  \label{eq:dYdx}
  \frac{dY_X}{dx} = - \sqrt{\frac{8\pi^2 g_*(T)}{45}}  \frac{m_X M_p}{x^2}
      \vev{\sigma \vrel} \left( Y_{X}^2-Y_{X,\eq}^2  \right), 
\end{align}
with $Y_{X,\eq} = n_{X,\eq}/s$ and $M_p=2.4\times 10^{18}~\GeV$ being the reduced Planck scale.   
Here, $g_*(T)$ is the number of relativistic degrees of freedom at the temperature $T$~\footnote{
  We assume that the effective degrees of freedom for the entropy density are
  the same as those for the energy density, which is valid if there is no
  additional thermal bath decoupled from the SM bath.  
  If all of the SM particles are in the bath, $g_*(T) = 106.75$.  
}.

Approximately, the asymptotic yield  is given by
\begin{align}
  \label{eq-solapp}
  \frac{1}{Y_X(\infty)} \simeq
  \sqrt{\frac{8\pi^2g_*(T_\fo)}{45}}\frac{M_p}{m_X}
  \left( \frac{m_X^2 a}{x_\fo} + \frac{3m_X^2 b}{x_\fo^2}
             + \frac{20m_X^2 c}{x_\fo^3} \right), 
\end{align}
where we have neglected the contribution $Y_X(x_\fo)^{-1}$ to the final inverse yield. 
Here, $x_\fo\simeq 20$ is the value of $x$ when $Y_X$ becomes much larger than $Y_{X,\eq}$.
Using this, the relic density in the $s$-wave and $d$-wave dominant cases is given by
\begin{align}
  \Omega_X h^2 =&\ \frac{h^2s_0}{\rho_c} m_X Y_X(\infty)
\\ \notag                   
  \simeq&\ 0.11 \times \left( \frac{106.75}{g_*(T_\fo)} \right)^\frac{1}{2}
  \left( \frac{m_X}{100~\GeV} \right)^2 \times 
  \begin{cases}
    \left( \dfrac{x_\fo}{20} \right)   \left( \dfrac{0.000015}{a m_X^2} \right)
    & \mathrm{s{-}wave} \\
\\
    \left( \dfrac{x_\fo}{20} \right)^3 \left( \dfrac{0.0003}{c m_X^2} \right) 
& \mathrm{d{-}wave} 
  \end{cases},
\end{align}
where ${h^2s_0}/{\rho_c} = 2.744\times 10^{8}~\GeV^{-1}$.

We numerically solve Eq.~\eqref{eq:dYdx} with the initial condition
$Y_X = Y_{X,\eq}$ at $x = x_i = 10$, before thermal freeze-out.
Figure~\ref{fig-lambda} shows the values of the Yukawa couplings required to reproduce
the observed DM density $\Omega_X h^2 = 0.12$~\cite{Planck:2018vyg}.
The left and right panels show the results for
$s_L = s_R = 0.1$ and $0.5$, respectively.
The blue lines show the $d$-wave dominated case with $\la_E = 0$,
while the red lines show the case with
$\la_V := \la_L = \la_E$, where the $s$-wave contribution is present.
The mass difference of the vector-like leptons is fixed at
$\Delta m_E := m_{E_2}-m_{E_1} = 100~\GeV$.

We see that the portal Yukawa coupling must be larger than unity
in much of the parameter space.
In the $d$-wave dominated case with $\lambda_E=0$,
it can exceed the perturbative value $\sqrt{4\pi}$.
When $\lambda_L\lambda_E\neq 0$, the required coupling decreases
as the mixing angles increase, since the $s$-wave contribution becomes more important.
Thus, away from the co-annihilation region,
the observed relic density can be reproduced by portal Yukawa couplings of
$\mathcal{O}(1)$.

\subsection{DM and LHC constraints}

We summarize the constraints from DM detection experiments and the LHC searches.
See Refs.~\cite{Bai:2014osa,Kawamura:2020qxo} for details.

For the real scalar DM,
the DM-nucleon scattering occurs through the diphoton exchange at the 2-loop level~\cite{Kopp:2009et},
and thus the detection rate is well below the sensitivities
of current experiments~\cite{XENON:2023cxc,LZ:2022lsv,PandaX-4T:2021bab}.
If the DM dominantly couples to the electron,
the DM-electron scattering can be probed
by low-threshold experiments,
such as SENSEI~\cite{SENSEI:2020dpa,SENSEI:2023gie}, DAMIC~\cite{DAMIC:2019dcn},
DarkSide~\cite{DarkSide:2018ppu} and SuperCDMS~\cite{SuperCDMS:2018mne},   
for sub-GeV DM~\cite{Essig:2011nj,Essig:2015cda,Essig:2017kqs,Okawa:2020jea}.   
This is an interesting possibility, but is irrelevant for $m_X \gtrsim 100~\GeV$  studied in this work.

The indirect detection signals  
are qualitatively different depending on whether $\la_L\la_E = 0$ or not.
If $\lambda_L\lambda_E =0$,
the tree-level annihilation into leptons is $d$-wave suppressed and is negligible in the present Universe.
In this case, loop-induced annihilation processes such as
$XX\to \gamma\gamma,\gamma Z,ZZ,WW$ can provide the leading
indirect-detection signals.
As studied in Ref.~\cite{Kawamura:2020qxo},
the viable parameter space is partially covered by the constraints
from Fermi/H.E.S.S. observations~\cite{McDaniel:2023bju,HESS:2016mib,HESS:2018cbt},
and a large part of the remaining region could be probed
by future CTA observations~\cite{Pierre:2014tra,Silverwood:2014yza}.
If $\la_L\la_E \ne 0$, 
the signal is induced from the $XX \to \ell_i \ol{\ell}_i$ dominated by the $s$-wave contribution.
The annihilation cross section in the present Universe is then close to the thermal value
$\vev{\sigma v}_{\mathrm{th}} = \order{10^{-26}}~\mathrm{cm}^3\mathrm{s}^{-1}$.
The lower bound on the DM mass from gamma-ray searches using the
Fermi-LAT dwarf-spheroidal data is typically $\mathcal{O}(100)\,\GeV$
for thermal annihilation into $b\bar b$ or $\tau^+\tau^-$ final states~\cite{FermiLAT:2015att}. 
Although a dedicated recast for the present model is beyond the scope of this
work, this indicates that indirect detection can be relevant
for the $s$-wave dominated case.

The mediator vector-like leptons can be produced at the LHC.
The relevant process is
$pp\to E_a\overline{E}_a\to \ell\bar\ell+E_T^{\rm miss}$,
which gives the same final state topology as slepton searches.
Using the ATLAS Run-2 data~\cite{ATLAS:2019lff},
the lower limit on the vector-like lepton mass is about $900~\GeV~(800~\GeV)$
for the doublet (singlet) vector-like lepton~\cite{Kawamura:2020qxo},
when the SM lepton is $e$ or $\mu$.
The limit is expected to be weaker if the SM lepton is $\tau$~\cite{ATLAS:2024fub,CMS:2024gyw}.
It is also possible that the vector-like leptons are pair produced
at the future lepton colliders, such as the ILC~\cite{Mahmoud:2024sby},
and muon colliders~\cite{Jueid:2023zxx,Asadi:2024jiy,De:2026wqw}.
Explicit recasts of the current data and forecasts for future experiments are important 
but beyond the scope of this paper.

Altogether, the direct detection rate is below the current experimental sensitivities,
while indirect searches and LHC searches can probe the LPDM
for DM masses around the electroweak scale
and vector-like lepton masses around the TeV scale, respectively.

\subsection{Flavor problem}

The DM can couple to more than one flavor of the SM leptons
through the portal Yukawa couplings.
The portal couplings induce lepton flavor violation~\footnote{
  The lepton flavor violation in the LPDM model
  is studied in Refs.~\cite{Kile:2013ola,Kile:2014jea,Chen:2015jkt,Desai:2020alk,Acaroglu:2023cza,Belfatto:2025ids}.  
},
especially $\ell_i \to \ell_j\gamma$,
whose branching fraction is given by~\cite{Lavoura:2003xp},    
\begin{align}
  \label{eq:LFVwidth}
  \mathrm{Br}\left(\ell_i\to\ell_j \gamma\right)
  =&\
     \frac{\alpha_e m_{\ell_i}^5}{1024\pi^4 m_X^4\Gamma_{\ell_i}}
     \left( 1-\frac{m_{\ell_j}^2}{m_{\ell_i}^2} \right)^3 
  \\ \notag
&  \times \left(
  \abs{ \sum_{a=1}^2\left\{
        \left(y_L^{aj*} y_L^{ai} + \frac{m_{\ell_j}}{m_{\ell_i}} y_R^{aj*} y_R^{ai}\right) F(r_a) 
  + \frac{m_{E_a}}{m_{\ell_i}} y_L^{aj*} y_R^{ai}  G(r_a) \right\} }^2
     + \left(L\leftrightarrow R \right )
     \right), 
\end{align}
where the functions $F$ and $G$ are defined in Eq.~\eqref{eq-defHFG}. 

When $y_R^{ai} \propto \la_E = 0$ and $s_L = s_R=0$ for simplicity,
\begin{align}
  \mathrm{Br}\left(\mu\to e\gamma\right)
  \sim&\
        \frac{\alpha_e m_\mu^5}{1024\pi^4m_X^4 \Gamma_\mu} F(r_1)^2 \abs{y_{L}^{1e} y_L^{1\mu}}^2 
\\ \notag         
  \sim&\ 3.2\times 10^{-14} \times  
             \left(\frac{100~\GeV}{m_X}\right)^4
             \left(\frac{F(r_1) \abs{y_{L}^{1e} y_L^{1\mu}}}{10^{-6}}\right)^2,   
  \\
  \mathrm{Br}\left(\tau\to \ell_j \gamma\right)
  \sim&\
  \frac{\alpha_e m_\tau^5}{1024\pi^4m_X^4 \Gamma_\tau} 
  F(r_1)^2 \abs{y_L^{1j} y_L^{1\tau}}^2 
\\ \notag                                               
  \sim&\ 5.7\times 10^{-7} \times   \left(\frac{100~\GeV}{m_X}\right)^4
        \left(\frac{F(r_1)\abs{y_L^{1j}y_L^{1\tau}} }{10^{-2}}\right)^2, 
\end{align}
where $\ell_j = e,\mu$ and the daughter lepton mass is neglected. 
The 90\% C.L. upper bounds on the branching fractions are~\cite{MEGII:2023ltw,MEGII:2025new,Belle:2009hvk,Belle:2021ysv,BaBar:2009hkt,ParticleDataGroup:2026aaa},   
\begin{align} 
  \label{eq:LFVlimits}
  \mathrm{Br}\left(\mu\to e\gamma\right) < 1.5\times 10^{-13},
  \quad
  \mathrm{Br}\left(\tau\to e\gamma\right) < 3.3\times 10^{-8},
  \quad
  \mathrm{Br}\left(\tau\to \mu\gamma\right) < 4.2\times 10^{-8}.   
\end{align}
Since portal Yukawa couplings of $\mathcal{O}(1)$ are required to reproduce
the relic density away from the co-annihilation region, the couplings to more
than one SM lepton flavor are severely constrained.
Therefore, a viable LPDM setup requires an approximate
one-flavor dominance of the portal couplings.
If $\lambda_L\lambda_E\neq 0$,
the chirality enhanced term proportional to $m_{E_a}/m_{\ell_i}$
gives a stronger constraint on the flavor-violating
combination of portal couplings by roughly $m_{E_a}^2/m_{\ell_i}^2$.
In phenomenological studies, it is often assumed that the portal Yukawa couplings 
are non-zero only for one of the three flavors.

\section{Taming LPDM by a non-invertible selection rule} 
\label{sec-NISR}

The main purpose of this paper is to show that a non-invertible selection rule
can realize the one flavor dominance of the portal couplings.
As shown below, this cannot be achieved by ordinary invertible symmetries
once the large mixing angles in the PMNS matrix are required.

\subsection{Case of ordinary symmetry}

We first try to tame the LPDM by a $\mathbb{Z}_N$ or $U(1)$ symmetry.
Without loss of generality, we set the charge of $X$ to zero, $q_X=0$,
and assume that the portal Yukawa coupling labeled by $\lambda_L^1$ is nonzero~\footnote{
  In this section, the flavor label $1$ does not necessarily correspond to $e$.
  It may instead correspond to $\mu$ or $\tau$.
  The label $1$ denotes the lepton flavor that has the portal coupling to the DM.  
}. 
The flavor violation is absent if 
\begin{align}
  \label{eq-qellcond}
  q_{\ell}^1 + q_{\ol{L}} = 0,
  \quad
  q_{\ell}^{2,3} + q_{\ol{L}}  \ne 0,  
\end{align}
where the equality is understood modulo $N$.

To make all three charged leptons massive, at least one element of $y_e$
should be nonzero in each row and column.
Without loss of generality, we relabel the singlet leptons $e_i$
so that one nonzero element appears on each diagonal entry.
With this convention, the charge condition is
\begin{align}
  \label{eq-fullrank}
  q_{\ol{e}}^i + q_\ell^i - q_H = 0, 
\end{align}
for $i=1,2,3$.
From Eq.~\eqref{eq-qellcond}, one finds 
\begin{align}
  \label{eq-qsele}
  q_{\ol{e}}^{2,3} + q_\ell^1 - q_H \ne 0 , 
\end{align}
so the first column of the charged lepton mass matrix is $(\checkmark, 0,0)$,
where $\checkmark$ indicates that the element can be non-zero.
Applying the same argument to the other columns, the first elements are zero 
and so the charged mass matrix has the texture 
\begin{align}
  \label{eq-MeCase1}
  \Mcal_e =
  \begin{pmatrix}
    \checkmark & 0 & 0 \\
    0 & \checkmark & \checkmark \\
    0 & \checkmark & \checkmark \\
  \end{pmatrix}    
\end{align}
where the off-diagonal elements are non-zero only if $q_{\ell}^2=q_{\ell}^3$.
Thus, the first lepton is block-diagonalized
and there is no mixing with the other charged leptons. 
The conclusion is the same when we require $\la_E^i \propto \delta^{i1}$.

The large mixing angles of the first lepton $\ell_1$ in the PMNS matrix
should be induced from the neutrinos. 
An element $c_{ij}$ is non-zero if 
\begin{align}
  q_{\ell}^i + q_{\ell}^j +  2q_H = 0.  
\end{align}
It turns out that there are three possible textures,
\begin{align}
  \label{eq:Mncands}
  \Mcal_n =
  \begin{pmatrix}
    \checkmark & 0 & 0 \\
    0 & \checkmark & \checkmark \\
    0 & \checkmark & \checkmark \\
  \end{pmatrix},
 \quad 
  \begin{pmatrix}
    0 & \checkmark & \checkmark \\
    \checkmark & 0 & 0 \\
    \checkmark & 0 & 0 \\
  \end{pmatrix},    
\quad 
  \begin{pmatrix}
    0 & \checkmark & 0 \\
    \checkmark & 0 & 0 \\
    0 & 0 & \checkmark \\
  \end{pmatrix}.     
\end{align}
The first texture is obtained when $2q_\ell^1+2q_H=0$.
The flavor-conserving condition in Eq.~\eqref{eq-qsele} then implies
$q_\ell^1+q_\ell^{2,3}+2q_H\neq0$.
The $(2,3)$ element can be nonzero only if
$q_\ell^2+q_\ell^3+2q_H=0$.
For $\mathbb{Z}_N$ with even $N$, the $(i,i)$ element, with $i=2,3$,
can be nonzero only if
$q_\ell^i = q_\ell^1 + N/2$ modulo $N$.
Thus, there is no mixing between the first lepton and the others,
and the PMNS matrix cannot be reproduced.

The other two textures in Eq.~\eqref{eq:Mncands}  
are reproduced when $q_{\ell}^1 + q_{\ell}^2 + 2q_H = 0$.
From Eq.~\eqref{eq-qsele}, 
one finds $2q_{\ell}^1 + 2q_H\ne 0$ and
$2q_{\ell}^2 + 2q_H \ne 0$.
The texture is constrained to be the second one
when $q_{\ell}^2 = q_{\ell}^3$
and to be the third one when $q_{\ell}^2 \ne q_{\ell}^3$.
In the latter case, the $(3,3)$ element is nonzero only if
$2q_\ell^3 + 2q_H = 0$.
These two textures are diagonalized by a unitary matrix
\begin{align}
  U_n = U_{23} \times \frac{1}{\sqrt{2}}
  \begin{pmatrix}
    1 & 1 & 0 \\
   -1 & 1 & 0 \\ 
    0 & 0 & \sqrt{2}    
  \end{pmatrix}, 
\end{align}
where $U_{23}$ rotates the second and third flavor components
so that the $(1,3)$ element becomes zero for the second texture,
while $U_{23}$ is the identity matrix for the third texture.
Thus, there is no continuous mixing angle involving the first flavor
that can be fitted to the PMNS matrix.

One could realize the mixing by introducing appropriate spurion fields
while keeping the flavor-violating portal couplings forbidden.
This is always possible in principle, but it often looks artificial or ad hoc.
Phenomenologically, such spurions may also induce flavor violation
because they must couple to multiple flavors in order to generate the mixing. 
Therefore, it is preferable to consider another possibility
that simultaneously forbids flavor-violating portal couplings
and realizes the observed lepton flavor structure~\footnote{
This issue is addressed in an extra-dimensional setup in Ref.~\cite{Desai:2020alk}.
}.

\subsection{Case of non-invertible selection rule}

We now show that a non-invertible selection rule can forbid
flavor-violating portal couplings while allowing the large mixing angles
in the PMNS matrix.
The key difference from an ordinary invertible symmetry is that, 
after gauging the $\mathbb{Z}_2$ automorphism, 
fields are assigned to conjugacy classes
rather than to ordinary group charges, and the allowed interactions are
determined by the corresponding fusion rule.

\subsubsection{$\mathbb{Z}_2$ gauging of $\mathbb{Z}_N$ symmetry} 

We introduce a realization of non-invertible selection rules
by gauging the $\mathbb{Z}_2$ automorphism of a $\mathbb{Z}_N$
symmetry~\cite{Kobayashi:2024cvp}. 
We consider a $\mathbb{Z}_N$ symmetry whose elements are denoted by
\begin{align}
  g^k, \quad k = 0,1,\cdots,N-1.  
\end{align}
The product rule is given by 
\begin{align}
  g^{k_1} g^{k_2} = g^{k_1+k_2}, 
\end{align}
where the exponent is understood modulo $N$.
For a general operator
\begin{align}
  \label{eq-genOp}
  \mathcal{O} = \prod_{i=1}^m \phi_i,
\end{align}
where a field $\phi_i$ has a $\mathbb{Z}_N$ charge $k_i$,
the ordinary selection rule requires
\begin{align}
  \prod_i g^{k_i} = g^{\sum_i k_i} = g^0
  \quad \Longleftrightarrow \quad 
  \sum_i k_i \equiv 0 \quad \mathrm{mod}~N.   
\end{align}
As argued in the previous subsection, such an ordinary $\mathbb{Z}_N$
selection rule cannot forbid flavor-violating portal couplings consistently
with the PMNS mixing matrix.

We then gauge the $\mathbb{Z}_2$ automorphism of $\mathbb{Z}_N$.
The $\mathbb{Z}_2$ is composed of two elements $\{e,r\}$, 
which act on
the $\mathbb{Z}_N$ elements as
\begin{align}
  e g^k e^{-1} = g^k, \quad
  r g^k r^{-1} = g^{-k}.   
\end{align}
Here, $e$ acts trivially and $r$ acts as a reflection.
After gauging this automorphism, the relevant charges are the classes
\begin{align}
  \label{eq:defClass}
  [g^k] := \left\{ h g^k h^{-1} \mid h = e,r \right\}
  = \{g^k,g^{-k}\}, 
\end{align}
which will be assigned to the fields.
The fusion rule of the classes is given by
\begin{align}
  \label{eq:niprod}
  [g^{k_1}][g^{k_2}]
  =
  [g^{k_1+k_2}] \oplus [g^{-k_1+k_2}],
\end{align}
where the exponents are understood modulo $N$.
In addition to the ordinary product $g^{k_1+k_2}$,
the second term appears because $g^{-k_1}$ belongs to the same class as
$g^{k_1}$.

Based on these classes, the operator in Eq.~\eqref{eq-genOp} is allowed
if the fusion product contains the trivial class $[g^0]$, namely
\begin{align}
  \label{eq:NIselectionFull}
  \prod_i [g^{k_i}]
  \supset [g^0].
\end{align}
Equivalently, there exists a choice of signs $\sigma_i=\pm1$ such that
\begin{align}
  \sum_i \sigma_i k_i \equiv 0 \quad \mathrm{mod}~N.
\end{align}

\subsubsection{Example of $N=5$}

For illustration and the model building,
we show the fusion rule of the $\mathbb{Z}_2$ gauging of $\mathbb{Z}_5$,
whose generators are given by
\begin{align}
  g^0,\quad g^1,\quad g^2,\quad g^3 = g^{-2}, \quad g^4 = g^{-1}. 
\end{align}
After the $\mathbb{Z}_2$ gauging, there are three classes 
\begin{align}
  [g^0], \quad [g^1] = \{g^1, g^4\} , \quad [g^2] = \{g^2, g^3 \}.  
\end{align}
The fusion rule is
\begin{align}
  \label{eq:fusionZ5}
  [g^0] [g^a] = [g^a],
  \quad
  [g^1] [g^1] = [g^0] \oplus [g^2],
  \quad
  [g^1] [g^2] = [g^1] \oplus [g^2],
  \quad
  [g^2] [g^2] = [g^0] \oplus [g^1],
\end{align}
where $a=0,1,2$.
Note that $[g^a][g^a] \ne [g^0]$ for $a = 1,2$
and there is no inverse element for $[g^a]$ in a usual sense.

\subsubsection{Application to lepton portal DM}

Now we apply the non-invertible selection rule
based on the $\mathbb{Z}_2$ gauging of $\mathbb{Z}_5$ to the LPDM model.
We first consider the following assignment of the classes:
\begin{align}
  \label{eq:Z5assign}
  &   X = [g^0], \quad L = [g^1], \quad E  = [g^2],
  \\ \notag
  &   H = [g^2], \quad
      \vec{\ell} = \left([g^1], [g^2], [g^2] \right), \quad
      \vec{e}    = \left([g^2], [g^0], [g^0] \right).
\end{align}
With this assignment, the portal couplings are constrained as
\begin{align}
\label{eq-portalGauge}
  \la_L \sim&\ [g^0] [g^1]
  \begin{pmatrix}
    [g^1] \\ [g^2] \\ [g^2]
  \end{pmatrix}
  =
  \begin{pmatrix}
    [g^0] \oplus [g^2] \\
    [g^1] \oplus [g^2] \\
    [g^1] \oplus [g^2]
  \end{pmatrix}
  \sim
  \begin{pmatrix}
    \checkmark \\ 0 \\ 0
  \end{pmatrix},
  \\ \notag
  \la_E \sim&\ [g^0] [g^2]
  \begin{pmatrix}
    [g^2] \\ [g^0] \\ [g^0]
  \end{pmatrix}
  =
  \begin{pmatrix}
    [g^0] \oplus [g^1] \\
    [g^2] \\
    [g^2]
  \end{pmatrix}
  \sim
  \begin{pmatrix}
    \checkmark \\ 0 \\ 0
  \end{pmatrix}.
\end{align}
Thus, flavor-violating portal couplings are forbidden by the
non-invertible selection rule in the gauge basis.

The $d$-wave dominated scenario can also be realized. 
If $L=[g^0]$ while the other fields are assigned as in Eq.~\eqref{eq:Z5assign}, all elements of $\lambda_L$ are
forbidden and only $\lambda_E^1$ is allowed.  Similarly, if $E=[g^1]$,
all elements of $\lambda_E$ are forbidden and only $\lambda_L^1$ is allowed.
In either case, $\lambda_L\lambda_E=0$ is naturally realized by the
non-invertible selection rule.

The texture of the charged lepton Yukawa matrix is read from
\begin{align}
  \label{eq:YeTexture}
  Y_e \sim&\ [g^2]
  \begin{pmatrix}
    [g^2] \\ [g^0] \\ [g^0]
  \end{pmatrix}
  \begin{pmatrix}
    [g^1] & [g^2] & [g^2]
  \end{pmatrix}
  \notag \\
  \sim&\
  \begin{pmatrix}
    [g^0]\oplus [g^1] \oplus [g^2] &
    [g^1]\oplus [g^2] &
    [g^1]\oplus [g^2] \\
    [g^1]\oplus [g^2] &
    [g^0]\oplus [g^2] &
    [g^0]\oplus [g^2] \\
    [g^1]\oplus [g^2] &
    [g^0]\oplus [g^2] &
    [g^0]\oplus [g^2]
  \end{pmatrix}
  \notag  \\
  \sim&\
  \begin{pmatrix}
    \checkmark & 0 & 0 \\
    0 & \checkmark & \checkmark \\
    0 & \checkmark & \checkmark
  \end{pmatrix}.
\end{align}
This texture can have three non-zero singular values, and the first flavor is
block diagonalized from the other two.  The latter feature ensures that
the texture of the portal coupling in Eq.~\eqref{eq-portalGauge} is
preserved in the charged lepton mass basis~\footnote{
  $N=5$ is the minimal possibility for realizing the block-diagonal texture of
  $Y_e$, the absence of flavor violation in the charged lepton mass basis, and the PMNS mixing matrix simultaneously.
}.

The texture of the Weinberg operator is also determined by the same selection
rule.  Since $H=[g^2]$, the two Higgs fields give
\begin{align}
  [g^2][g^2] = [g^0]\oplus [g^1].
\end{align}
Therefore,
\begin{align}
  c_n
  \sim&\
  \left( [g^0] \oplus [g^1] \right)
  \begin{pmatrix}
   [g^0]\oplus [g^2] &
   [g^1]\oplus [g^2] &
   [g^1]\oplus [g^2] \\
   [g^1]\oplus [g^2] &
   [g^0]\oplus [g^1] &
   [g^0]\oplus [g^1] \\
   [g^1]\oplus [g^2] &
   [g^0]\oplus [g^1] &
   [g^0]\oplus [g^1]
  \end{pmatrix}
  \notag \\
  \sim&\
  \begin{pmatrix}
    \checkmark & \checkmark & \checkmark \\
    \checkmark & \checkmark & \checkmark \\
    \checkmark & \checkmark & \checkmark
  \end{pmatrix}.
\end{align}
Each entry contains the trivial class $[g^0]$ after multiplication by
$[g^0]\oplus[g^1]$.  We therefore find that all elements of the Weinberg
operator are allowed, and the model can realize a general PMNS matrix through
the neutrino diagonalization matrix $U_n$.

Altogether, we have found an assignment of the classes of the
$\mathbb{Z}_2$ gauging of $\mathbb{Z}_5$ that forbids flavor violation
induced by the portal couplings in the charged lepton mass basis.
In this setup, the PMNS matrix can be realized by the diagonalization matrix
of the neutrino sector, $U_n$.
Thus, the lepton portal DM can be tamed by a non-invertible selection rule,
which cannot be achieved by ordinary symmetries.

\subsection{Flavor leakage in type-I seesaw}

Since the non-invertible selection rule is not protected by an ordinary
invertible symmetry, quantum corrections may violate the selection rule.
For illustration, we consider the type-I seesaw mechanism as a UV completion
of the Weinberg operator and discuss possible flavor violation induced by
renormalization group (RG) effects.
See Refs.~\cite{Suzuki:2025oov,Suzuki:2025bxg,Suzuki:2025kxz}
for more general discussions on violations of non-invertible selection rules.

\subsubsection{Type-I seesaw model}

We introduce three flavors of right-handed neutrinos,
whose masses and Yukawa couplings are given by
\begin{align}
  \label{eq:Lrhn}
  - \Lcal_n =
  \frac{1}{2} M_n^{ij} \ol{n}_{i} P_L n^{c}_j
  + y_n^{ij} \ol{n}_i P_L \ell_j \cdot H
  + h.c.
\end{align}
After integrating out the right-handed neutrinos,
the coefficients of the Weinberg operator are matched to
\begin{align}
  \frac{c_n^{ij}}{\Lambda_n}
  =
  \sum_{\alpha,\beta=1}^3
  y_n^{\alpha i}
  \left(M_n^{-1}\right)_{\alpha\beta}
  y_n^{\beta j}.
\end{align}

If we assign $[g^1]$ to all of the right-handed neutrinos,
all elements of the Majorana mass matrix $M_n$ and the Yukawa matrix $y_n$
are nonzero.
By contrast, if we assign other classes, namely $[g^0]$ or $[g^2]$,
the selection rules predict textures with zeros in these matrices,
which would lead to non-trivial textures in the coefficient $c_n^{ij}$ at tree-level.
This discrepancy between the UV completion and the effective theory
may be understood in terms of loop effects, such as RG and threshold effects.
The additional hierarchical structure may arise from the additional selection
rule in the UV completion.
This feature is interesting to study, but is beyond the scope of this paper.
We consider the trivial case with
$n = ([g^1], [g^1], [g^1])$,
so that all elements of $M_n$ and $y_n$ are nonzero.

\subsubsection{Flavor leakage}

We discuss possible flavor leakage of the portal coupling induced by RG effects.
At the 1-loop level, the beta functions of the portal couplings are given by~\cite{Machacek:1983tz,Machacek:1983fi,Luo:2002ti}
\begin{align}
  \label{eq:betaXRGE}
  16\pi^2 \beta_{\la_L}
  =&\ 
     \left(
     7 \la_L \la_L^\dagger +  2 \la_E^\dag \la_E
     + \frac{1}{2} y_e^T y_e^*
     + \frac{1}{2} y_n^T y_n^*
     + \frac{1}{2}\abs{\ol{\ka}}^2
     -\frac{9}{10} g_1^2 -\frac{9}{2} g_2^2
     \right) \la_L
     + 2 \ol{\ka} \la_E^\dag y_e,
\notag       \\
  16\pi^2 \beta_{\la_E}
  =&\
     \left(
     5 \la_E\la_E^\dag 
     + 4 \la_L^\dag \la_L
     + y_e y_e^\dag
     + \abs{\ol{\kappa}}^2
     -\frac{18}{5} g_1^2
     \right) \la_E
     + 4y_e \la_L^* \ol{\kappa}.
\end{align}
Here, we omit the flavor indices,
and treat $y_{e,n}$ as $3\times 3$ matrices
and $\la_{L,E}$ as three-component column vectors.

With the texture in our model,
flavor mixing can be induced by the neutrino Yukawa coupling $y_n$.
Thus, assuming that the non-invertible selection rule is exact at a scale
$\Lambda$, the leading leakage of the portal coupling at the seesaw scale
$\Lambda_n$ is estimated as
\begin{align}
  \label{eq:portalLeak}
  \la_L(\Lambda_n) \sim&\
  \left(
     1 + \frac{y_n^T y_n^*}{32\pi^2}
     \log \frac{\Lambda_n}{\Lambda}
  \right) \la_L(\Lambda).
\end{align}
Since $y_n$ appears only in $\beta_{\la_L}$ at the 1-loop level,
the flavor leakage of $\la_E$ is induced only at the 2-loop level.
Thus, flavor violation by the portal couplings is correlated with the
neutrino Yukawa coupling in the type-I seesaw model.

Without loss of generality, we start from the basis
in which the charged lepton Yukawa matrix is diagonal.
For the neutrino sector, we parametrize the matrices as
\begin{align}
  M_n = \mathrm{diag}\left( M_{1}, M_2, M_3 \right),
  \quad
  y_n v_H = u_R^\dag d_n u_L,
  \quad
  d_n :=  \mathrm{diag}\left( d_{1}, d_2, d_3 \right),
\end{align}
where $u_{L,R}$ are unitary matrices.
In this case, the neutrino mass matrix is diagonalized as
\begin{align}
  \label{eq:ndiag}
 M_\nu =
 U_{\mathrm{PMNS}}^T u_L^T d_n u_R^* M_n^{-1}
 u_R^\dag d_n u_L U_{\mathrm{PMNS}}
 =
 \mathrm{diag}\left( m_{\nu_1}, m_{\nu_2}, m_{\nu_3} \right).
\end{align}
In a special case with $u_R = \mathbf{1}_3$,
\begin{align}
  \label{eq:simplecase1}
  u_L = U_{\mathrm{PMNS}}^\dag,
  \quad
  M_\nu =
  \diagp{d_{1}^2/M_{1}, d_{2}^2/M_{2}, d_{3}^2/M_{3}}.
\end{align}
In another special case with $u_L = \mathbf{1}_3$
and $d_n = m_n \mathbf{1}_3$,
\begin{align}
  \label{eq:simplecase2}
  u_R = U_{\mathrm{PMNS}},
  \quad
  M_\nu =
  m_{n}^2
  \diagp{1/M_{1}, 1/M_{2}, 1/M_{3}}.
\end{align}
These two cases can be consistent with the neutrino oscillation data.
The flavor leakage is directly determined by the PMNS matrix in the first case,
while it is absent in the second case.

In the first case, further assuming $M_n = m_N \mathbf{1}_3$,
the size of the flavor-violating portal coupling is estimated as
\begin{align}
  \frac{\la_L^{i}}{\la_L^1}
  \sim&\
  \frac{\left[y_n^T y_n^{*}\right]_{i1}}{32\pi^2}
  \log \frac{\Lambda_n}{\Lambda}
  \sim 
  \frac{m_N}{32\pi^2 v_H^2}
  \left[
  U^*_{\mathrm{PMNS}} M_\nu U^T_{\mathrm{PMNS}}
  \right]_{i1}
  \log \frac{\Lambda_n}{\Lambda}
  \notag \\
  \sim&\
   0.005 \times
        \left(\frac{m_N}{10^{14}~\GeV}\right)
        \left(\frac{\log (\Lambda/\Lambda_n)}{10}\right)
        \left(\frac{
        \left[ U^*_{\mathrm{PMNS}} M_\nu U^T_{\mathrm{PMNS}} \right]_{i1}}
        {0.05~\mathrm{eV}}\right).
\end{align}
Thus, the leakage could be probed by lepton flavor violation, and its pattern
is connected to the Yukawa couplings of the right-handed neutrinos.

\section{Phenomenology of leptonic observables}
\label{sec-pheno}

Before closing this paper, we discuss the phenomenology of the
flavor conserving portal coupling.
We study the anomalous magnetic moments of the leptons
and the flavor universality of the $Z$ boson couplings.
The 1-loop amplitudes relevant to these observables are calculated in
App.~\ref{app-formula}.

\subsection{Anomalous magnetic moment}

\begin{figure}[t] 
  \centering
  \begin{minipage}{0.48\textwidth}
    \centering
  \includegraphics[width=0.90\linewidth]{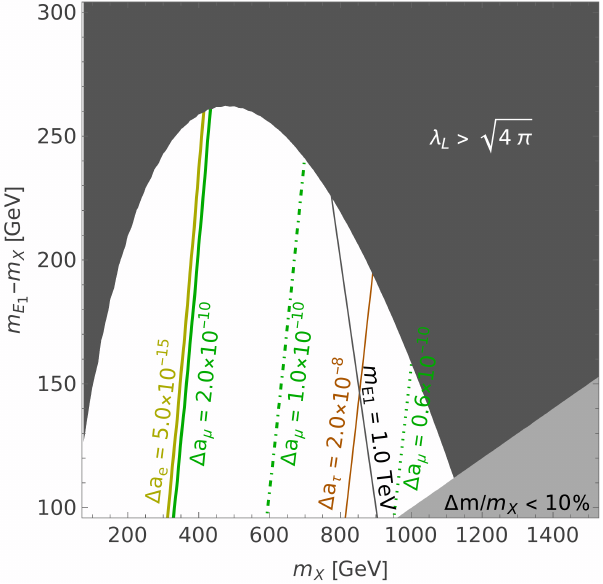}    
  \end{minipage}
  \begin{minipage}{0.48\textwidth}
    \centering    
  \includegraphics[width=0.90\linewidth]{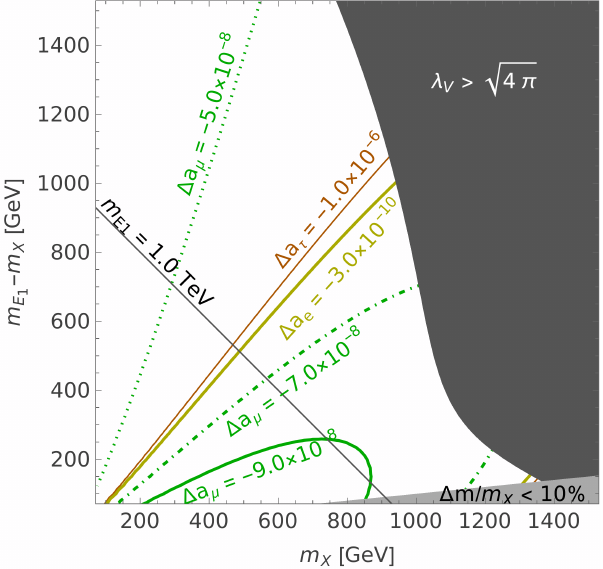}    
  \end{minipage}
  \caption{\label{fig_gm2}
The size of the anomalous magnetic moments of the SM leptons
when $\la_E = 0$ (left) and $\la_L = \la_E$ (right).
The Yukawa couplings are fixed to reproduce the observed DM relic density,
with $s_L = s_R = 0.5$ and $\Delta m_E = 100~\GeV$.
The Yukawa coupling exceeds the perturbative limit $\sqrt{4\pi}$
in the dark gray region, while the mass difference between the portal lepton
and the DM is less than 10\% of the DM mass in the light gray region,
where co-annihilation becomes relevant.
  }
\end{figure}

As also derived in Ref.~\cite{Kawamura:2020qxo},
the contribution to the anomalous magnetic moment
$\Delta a_\ell := (g-2)_\ell/2$ is given by 
\begin{align*}
  \Delta a_\ell
  =&\ \frac{m_\ell^2}{8\pi^2m_X^2} \left[
  \left( \abs{\la_Lc_R}^2  + \abs{\la_E s_L}^2  \right) F\left( r_1 \right)
  + \left( \abs{\la_Ls_R}^2    + \abs{\la_Ec_L}^2 \right) F\left( r_2 \right)
 \right.                    
\\ \notag                     
   &\ \left. \hspace{2.5cm}
     - \frac{m_{E_1}}{m_\ell}
     \mathrm{Re}\left( s_L^* c_R \la_L \la_E  \right) G\left( r_1 \right)
     + \frac{m_{E_2}}{m_\ell}
     \mathrm{Re}\left( c_L s_R^* \la_L \la_E  \right) G\left( r_2 \right)  
     \right],  
\end{align*}
where the functions $F(r)$ and $G(r)$ are the same as for $\ell_i \to \ell_j \gamma$.  
As for the DM annihilation,
$\Delta a_\ell$ becomes large if $\la_L \la_E \ne 0$ due to the chiral enhancement inside the loop.

For $\la_E = 0$, where the $d$-wave dominates the annihilation, 
$\Delta a_\ell$ can be directly related
to the DM density through the approximate solution of the Boltzmann equation~\eqref{eq-solapp} as  
\begin{align}
  \Delta a_\ell
  \sim&\ \frac{1}{8\pi^2}
  \left( \frac{45}{8g_*(T_\fo)} \right)^{\frac{1}{4}}
  \sqrt{\frac{s_0}{\Omega_X \rho_c M_p}}  
  \frac{\sqrt{3} m_\ell^2 x_\fo^{3/2} }{m_X}
  \frac{c_R^2 F(r_1)+ \abs{s_R}^2 F(r_2)}
        {\sqrt{f(r_n)^2+ \left( c_R^2 f(r_1) + \abs{s_R}^2 f(r_2) \right)^2}}
  \\ \notag
  \sim&\ 6.5\times 10^{-10} \times
        \left( \frac{m_\ell}{m_\mu} \right)^2
        \left( \frac{x_\fo}{20} \right)^{\frac{3}{2}}
        \left( \frac{500~\GeV}{m_X} \right)
        \frac{c_R^2 F(r_1)+ \abs{s_R}^2 F(r_2)}
        {\sqrt{f(r_n)^2+ \left( c_R^2 f(r_1) + \abs{s_R}^2 f(r_2) \right)^2}},         
\end{align}
where
\begin{align}
  f(r) := \frac{1}{(1+r)^2}.  
\end{align}
The typical deviation of the muon anomalous magnetic moment
is an order of magnitude smaller than the current sensitivity
$\Delta a_\mu \sim \order{10^{-9}}$.
It is also smaller than the current sensitivities
of $\Delta a_e \sim \order{10^{-13}}$ and $\Delta a_\tau \sim \order{10^{-3}}$. 
Thus, the anomalous magnetic moment does not provide a very sensitive probe
of the LPDM if $\la_L = 0$ or $\la_E = 0$.

If $\la_L\la_E \ne 0$, the annihilation is dominated by the $s$-wave,
and $\Delta a_\ell$ is approximately given by
\begin{align}
  \label{eq:alDMLR}
  \Delta a_\ell \sim&\ -\frac{m_\ell}{8\pi^2 }
                     \left( \frac{45}{8g_*} \right)^{\frac{1}{4}}
                       \sqrt{\frac{x_\fo s_0}{\Omega \rho_c M_p} }
  \frac{\mathrm{Re}(s_Lc_R)\sqrt{r_1}G(r_1)-\mathrm{Re}(c_Ls_R) \sqrt{r_2}G(r_2)}
  {\abs{s_Lc_R\sqrt{r_1 f(r_1)}-c_Ls_R \sqrt{r_2f(r_2)}}}
  \\ \notag
  \sim&\ -8.9\times 10^{-8} \times
        \left( \frac{m_\ell}{m_\mu} \right) 
        \left( \frac{x_\fo}{20} \right)^{\frac{1}{2}}
  \frac{\mathrm{Re}(s_Lc_R)\sqrt{r_1}G(r_1)-\mathrm{Re}(c_Ls_R) \sqrt{r_2}G(r_2)}
  {\abs{s_Lc_R\sqrt{r_1 f(r_1)}-c_Ls_R \sqrt{r_2f(r_2)}}}. 
\end{align}
This is an order of magnitude larger than the sensitivity
$\Delta a_\mu \sim \order{10^{-9}}$,
and thus the muon-philic case is excluded in the absence of co-annihilation.  
Similarly, the electron-philic case is also excluded
since it predicts $\Delta a_e \sim \order{10^{-10}}$.
By contrast, the tau-philic case predicts $\Delta a_\tau \sim \order{10^{-6}}$,
which is much smaller than the current sensitivity.

Figure~\ref{fig_gm2} shows the size of the anomalous magnetic moments
of the SM leptons when $\la_E = 0$ (left) and $\la_L = \la_E$ (right).
The Yukawa couplings are fixed to reproduce the observed DM relic density,
with $s_L = s_R = 0.5$ and $\Delta m_E = 100~\GeV$.
The Yukawa coupling exceeds the perturbative limit $\sqrt{4\pi}$
in the dark gray region, while the mass difference between the portal lepton
and the DM is less than 10\% of the DM mass in the light gray region,
where co-annihilation becomes relevant.
The results of the numerical analysis agree with the order-of-magnitude
estimates above, and the electron- and muon-philic cases with
$\la_L\la_E \ne 0$ are excluded outside the co-annihilation region.

Altogether, the lepton anomalous magnetic moment is directly correlated with
the annihilation rate that determines the relic density of DM.
The electron- and muon-philic DM are excluded by the current constraints
if $\la_L \la_E \ne 0$ and co-annihilation is absent.
If $\la_L=0$ or $\la_E = 0$, $\Delta a_e$ and $\Delta a_\mu$
are one order of magnitude smaller than the current sensitivities,
and they could be probed by future improvements.
The tau-philic case predicts $\Delta a_\tau$ much smaller than the current
sensitivity even if $\la_L \la_E \ne 0$.

\subsection{$Z$ boson flavor universality}

\begin{figure}[t] 
  \centering
  \includegraphics[width=0.55\linewidth]{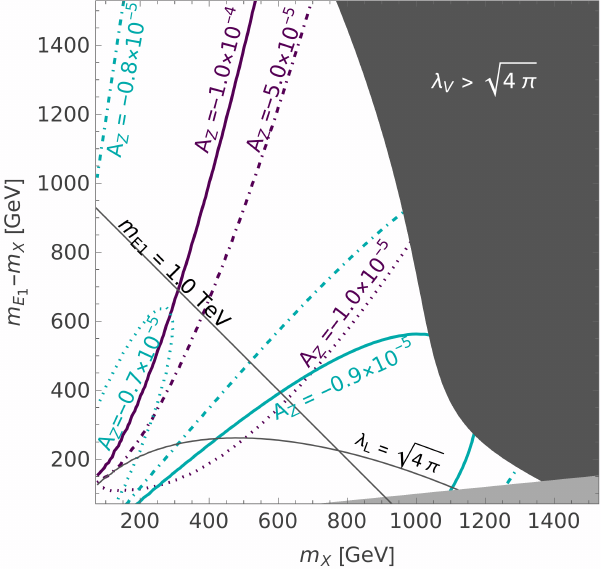}
  \caption{\label{fig-ZNLFU}
 Values of the lepton flavor non-universality in $Z$ boson decays,
$A_Z$.
The purple lines show the case with $\la_E = 0$,
while the cyan lines show the case with $\la_L = \la_E$.
The Yukawa couplings are fixed to reproduce the observed DM relic density,
with $s_L = s_R = 0.5$ and $\Delta m_E = 100~\GeV$.
For $\la_E = 0$, the Yukawa coupling $\la_L$ becomes non-perturbative
above the gray line, while in the case $\la_L = \la_E = \la_V$,
the region with $\la_V > \sqrt{4\pi}$ is shown in dark gray.
  }
\end{figure}

The sizable portal Yukawa couplings may modify the $Z$ boson couplings through loop corrections.
The 1-loop corrections to the $Z$ boson couplings are given by 
\begin{align*}
  \hat{g}_L^{ij}
  =&\ g_L^e \delta^{ij} + \delta g_L^{ij}
   = \frac{g}{c_W} 
               \left( -\frac{1}{2} + s_W^2 \right)\delta^{ij}
               + \frac{\la_L^{i*}\la_L^j}{32\pi^2} F_L,  
\\
  \hat{g}_R^{ij}
  =&\ g_R^e \delta^{ij} + \delta g_R^{ij}
   = \frac{g}{c_W}
                s_W^2\delta^{ij}
                + \frac{\la_E^i\la_E^{j*}}{32\pi^2}
                   F_R,  
\end{align*}
where $F_L$ and $F_R$ are defined in Eq.~\eqref{eq-defFLFR}.
As an order of magnitude estimate,
\begin{align}
  \label{eq-ZlfnuEst}
 \frac{\delta g_L^{11}}{g_L^e}
 \sim&
 \frac{\abs{\la_L^1}^2}{32\pi^2} \frac{m_Z^2}{m_L^2}
 \sim 2.6\times 10^{-5} \times 
 \abs{\la_L^1}^2
 \left( \frac{1~\TeV}{m_L} \right)^2. 
\end{align}
The partial decay width of the $Z$ boson into charged leptons is given by
\begin{align*}
  \Gamma(Z\to \ol{e}_ie_i)
   =&\ \frac{m_Z}{24\pi}
   \left( \abs{\hat{g}_L^{ii}}^2 + \abs{\hat{g}_R^{ii}}^2  \right)
\\
   \simeq&\ \frac{m_Z}{24\pi}
   \Bigl[
          \abs{{g}_L^e}^2 + \abs{{g}_R^e}^2
          + 2\mathrm{Re}\left(
          g_L^e \delta g_L^{ii} + g_R^{e} \delta g_R^{ii}
          \right) 
         \Bigr],                                   
\end{align*}
where the lepton masses are neglected.
Note that flavor-violating $Z$ decays are strongly suppressed by
$\order{(\delta g_{L,R})^2}$.

We define the lepton flavor non-universality in $Z$ boson decays as 
\begin{align}
  A_{Z}
  :=&\ \frac{\Gamma(Z\to \ol{e}_i e_i) - \Gamma(Z\to \ol{e}_j e_j)}
      {\Gamma(Z\to \ol{e}_i e_i) + \Gamma(Z\to \ol{e}_j e_j)}
  \simeq 
  \frac{\mathrm{Re}\left( g_L^e \delta g^{ii}_L
  + g_R^e \delta g_R^{ii} \right)}
       {\abs{g_L^e}^2+\abs{g_R^e}^2}
  + \order{(\delta g_{L,R})^2},  
\end{align}
where we assume that only the $i$-th lepton couples to the DM in the second
equality.  
Figure~\ref{fig-ZNLFU} shows the values of the lepton flavor non-universality
in $Z$ boson decays, $A_Z$.
The purple lines show the case with $\la_E = 0$,
while the cyan lines show the case with $\la_L = \la_E$.
The Yukawa couplings are fixed to reproduce the observed DM relic density,
with $s_L = s_R = 0.5$ and $\Delta m_E = 100~\GeV$.
We see that the non-universality is $\order{10^{-5}}$ as expected from Eq.~\eqref{eq-ZlfnuEst}.

The current experimental data of the leptonic decay modes are~\cite{ParticleDataGroup:2026aaa}
\begin{align*}
  \br{Z}{e^+e^-} =&\ 3.3632\pm 0.0043~\%,
\\ \notag 
  \br{Z}{\mu^+\mu^-} =&\ 3.3662\pm 0.0066~\%,
\\ \notag 
  \br{Z}{\tau^+\tau^-} =&\ 3.3696\pm 0.0083~\%. 
\end{align*}
Hence, the current uncertainty of the lepton flavor non-universality is
$\order{10^{-3}}$, 
while the LPDM predicts non-universality of
$\order{10^{-5}}$ from Eq.~\eqref{eq-ZlfnuEst}.
The relative uncertainty of $\order{10^{-5}}$ is expected at FCC-ee for
$R_\ell^Z:=\Gamma(Z\to\mathrm{had})/\Gamma(Z\to\ell^+\ell^-)$
and for ratios of leptonic $Z$ widths of different flavors~\cite{FCC:2018evy,FCC:2018byv}. 
A comparable Tera-$Z$ program is also envisaged at
CEPC~\cite{CEPCStudyGroup:2018ghi}. 
Thus, the lepton flavor non-universality in $Z$ boson decays provides
a promising probe of the flavor-conserving LPDM.

\section{Conclusion}
\label{sec-concl}

In this work, we have proposed a way to suppress flavor violation
in lepton portal DM (LPDM) consistently with the PMNS mixing angles.
We have shown that this cannot be achieved by ordinary symmetries
such as $\mathbb{Z}_N$ and $U(1)$.
We then presented an explicit example based on the $\mathbb{Z}_2$ gauging
of a $\mathbb{Z}_5$ symmetry, where the assignment of the classes is
summarized in Eq.~\eqref{eq:Z5assign}.
This is the minimal example that realizes suppressed flavor violation
in the mass basis of the SM leptons.

We also discussed possible flavor violation induced by RG effects
when the effective neutrino mass operator is generated by right-handed
neutrinos in the type-I seesaw model.
Flavor violation in the portal coupling $\la_L$ is induced at the 1-loop
level through the neutrino Yukawa matrix $y_n$.
Thus, the Yukawa matrix $y_n$ of the type-I seesaw model could be probed
through the pattern of lepton flavor violation.

The LPDM can also be probed by precision measurements of lepton observables.
The electron- and muon-philic cases with $\la_L\la_E \ne 0$
are excluded by the current data on the anomalous magnetic moments,
while the predicted deviations are about one order of magnitude smaller
than the current experimental sensitivities if $\la_L = 0$ or $\la_E=0$,
where there is no chiral enhancement.
The tau-philic LPDM is free from these constraints due to the current
experimental uncertainty.
An inevitable deviation also arises in the $Z$ boson couplings, inducing
lepton flavor non-universality.
The LPDM predicts deviations of $\order{10^{-5}}$, which could be probed by
future $Z$ factories.

\section*{Acknowledgement}

The work of J.K. is supported in part by JSPS KAKENHI Grant Number 25K00222.

\appendix

\section{Calculation of 1-loop observables}
\label{app-formula}

We calculate the 1-loop corrections to the gauge couplings through
the portal Yukawa couplings.
We consider internal fermions $\psi_a$ that couple to a vector boson $V_\mu$
and a real scalar $S$ as  
\begin{align}
  \label{eq:genint}
  \Lcal_{\mathrm{int}}
  = V_\mu \ol{\psi}_a \gamma^\mu \left(g^{ab}_L P_L + g_R^{ab} P_R\right) \psi_b
  - S \ol{\psi}_a \left(y^{ai}_L P_L + y^{ai}_R P_R \right) e_i
     - S \ol{e}_i \left(y^{ai*}_R P_L + y^{ai*}_L P_R \right) \psi_a.    
\end{align}
The external fermion masses are denoted by $m_i$ and $m_j$,
and the internal fermion masses are denoted by $m_a$.
The gauge boson and scalar masses are denoted by $m_V$ and $m_S$, respectively.
We neglect terms of $\order{m_i^2}$ and $\order{m_j^2}$ in the following.
We work in dimensional regularization with $d=4-2\eps$.

\subsection{Wave function renormalization}

We consider the general fermion Lagrangian in the mass basis,
\begin{align}
\label{eq-bareL}
 \Lcal = \ol{\psi}_i (i\dsdel - m_i ) \psi_i + \Lcal_{\mathrm{int}}, 
\end{align}
where the interactions may violate flavor.
We define the renormalized fields $\hat{\psi}_i$ and masses $\hat m_i$ by
\begin{align}
 \psi_i =: 
  \left(\delta_{ij} + \frac{1}{2} Z_L^{ij} P_L + \frac{1}{2} Z_R^{ij} P_R \right) 
  \hat{\psi}_j,
\quad
  \ol{\psi}_i =: \ol{\hat{\psi}}_j
    \left(\delta_{ji} + \frac{1}{2} \ol{Z}_R^{ji} P_L + \frac{1}{2} \ol{Z}_L^{ji} P_R \right),   
\end{align}
and
\begin{align*}
  m_i =: \hat{m}_i + \delta m_i.    
\end{align*}
In the on-shell scheme, the renormalization constants are given by 
\begin{align*}
  Z_X^{ii}
  =&\ -\frac{1}{32\pi^2}\sum_a \left[
     \left( \frac{1}{\eta} +B_1(m_i^2) +m_i^2 B^\pr_{1}(m_i^2) \right) \yy{X}{X}
     + m_i^2 B_{1}^\pr(m_i^2) \yy{\ol{X}}{\ol{X}} \right.
\\ \notag      
   &\ \left. \hspace{3.0cm}
     -\frac{2m_a}{m_i}\left( \frac{1}{\eta} + B_0(m_i^2) -m_i^2 B_{0}^\pr (m_i^2) \right)
     \yy{\ol{X}}{X}
     +2m_a m_i B_{0}^\pr (m_i^2) \yy{X}{\ol{X}}
     \right],
  \\ \notag
  \ol{Z}_X^{ii}
  =&\ -\frac{1}{32\pi^2}\sum_a\left[
     \left( \frac{1}{\eta} +B_1(m_i^2) +m_i^2 B^\pr_{1}(m_i^2) \right) \yy{X}{X}
     + m_i^2 B_{1}^\pr(m_i^2) \yy{\ol{X}}{\ol{X}} \right.
\\ \notag      
   &\ \left. \hspace{3.0cm}
     +\frac{2m_a}{m_i}\left( \frac{1}{\eta} + B_0(m_i^2) +m_i^2 B_{0}^\pr (m_i^2) \right)
     \yy{\ol{X}}{X}
     +2m_a m_i B_{0}^\pr (m_i^2) \yy{X}{\ol{X}}
     \right],   
\end{align*}
for the diagonal elements, and 
\begin{align*}
  \frac{1}{2} Z^{ij}_X
  =&\ \frac{1}{32\pi^2(m_i^2-m_j^2)}\sum_a\Biggl[
       m_j\left( \frac{1}{\eta} +B_{1}(m_j^2) \right)
     \left( m_j\yy{X}{X} +m_i \yy{\ol{X}}{\ol{X}} \right)
\\ \notag   
   &\  \hspace{4.0cm}
     +2m_a\left( \frac{1}{\eta}+ B_0(m_j^2)\right) \left( m_i \yy{\ol{X}}{X}+m_j \yy{X}{\ol{X}} \right) 
     \Biggr],
\\      
  \frac{1}{2}\ol{Z}^{ij}_X
  =&\ \frac{-1}{32\pi^2(m_i^2-m_j^2)}\sum_a\Biggl[
       m_i\left( \frac{1}{\eta} +B_{1}(m_i^2) \right)
     \left( m_i\yy{X}{X} +m_j \yy{\ol{X}}{\ol{X}} \right)
\\ \notag   
   &\  \hspace{4.0cm}
     +2m_a\left( \frac{1}{\eta}+ B_0(m_i^2)\right) \left( m_i \yy{\ol{X}}{X}+m_j \yy{X}{\ol{X}} \right)
     \Biggr],     
\end{align*}
for the off-diagonal elements $i\ne j$ with $X=L,R$.

The loop functions are defined as 
\begin{align*}
  B_0(p^2) := \int^1_0 dx (-1) \log \Delta_2(p^2),
  \quad
  B_1(p^2) := \int^1_0 dx (-2x) \log \Delta_2(p^2),  
\end{align*}
where
\begin{align*}
  \Delta_2(p^2) = x + (1-x) r_a - x (1-x) r_p,  
\end{align*}
with $r_a := m_a^2/m_S^2$ and $r_p := p^2/m_S^2$.
Here and in the following, 
$B_{0,1}(p^2)$ denote functions of $r_a$ and $r_p$. 
We also define
\begin{align*}
  B^\pr_{0}(p^2) := \left. \frac{\partial B_0(q^2)}{\partial q^2} \right|_{q^2 = p^2},
  \quad 
  B^\pr_{1}(p^2) := \left. \frac{\partial B_1(q^2)}{\partial q^2} \right|_{q^2 = p^2}. 
\end{align*}
The combinations of the Yukawa coupling constants are defined by
\begin{align*}
  y_{X}^\dag y_Y = y^{ai*}_{X} y^{aj}_Y,   
\end{align*}
where $\ol{L}=R$ and $\ol{R}=L$.

\subsection{Vertex correction}

We calculate the amplitude of $V_\mu(q) \to \ol{e}_i(p) e_j(k)$,
where the momentum $q^\mu$ is incoming and $p^\mu,~k^\mu$ are outgoing. 
The irreducible 1-loop amplitude is given by~\cite{Kawamura:2026ivz} 
\begin{align}
\delta \Mcal^\mu 
  =&\ \frac{1}{32\pi^2 m_S^2}\ol{e}_j \sum_{a,b} \Biggl[
     \gamma^\mu \Biggl(
      \left\{ m_S^2 \left(\frac{1}{\eta}-1-2C_{00} \right)  
     - q^2 C_{kp}
     \right\}  \ygy{L}{R}{L}
  +  m_am_b C_0 \ygy{L}{L}{L}
\\ \notag      
 &\  \hspace{2.5cm} 
 + m_a m_i (C_0-C_p) \ygy{L}{L}{R}
 - m_a m_j C_k  \ygy{R}{R}{L} 
\\ \notag      
 &\  \hspace{2.5cm} 
 + m_b m_i C_p \ygy{L}{R}{R}
 + m_b m_j (C_0+C_k) \ygy{R}{L}{L} 
     \Biggr)
\\ \notag   
&\  + 2p^\mu \Biggl(
  m_b C_p \ygy{R}{L}{L} + m_i (C_p-C_{pp}) \ygy{R}{L}{R}  + m_j C_{kp} \ygy{L}{R}{L} 
  \Biggr)        
\\ \notag    
&\  + 2k^\mu \Biggl(
   m_a C_k \ygy{R}{R}{L} - m_i C_{kp} \ygy{R}{L}{R}  + m_j (C_{k}+C_{kk}) \ygy{L}{R}{L} 
  \Biggr)
  \Biggr] P_L e_i + (L\leftrightarrow R),
\end{align}
where
\begin{align}
  \label{eq:defCs}
  C_0 :=&\ \int^{1}_0 dy 2y \int^1_0 dx \frac{1}{\Delta_3},
          \quad&
  C_{00} :=&\ \int^{1}_0 dy 2y  \int^1_0 dx\; \frac{1}{2} \log{\Delta_3},
  \\ \notag
  C_k :=&\  \int^{1}_0 dy 2y \int^1_0 dx  \frac{-(1-y)}{\Delta_3},
\quad& 
  C_p :=&\ \int^{1}_0 dy 2y \int^1_0 dx \frac{y(1-x)}{\Delta_3},
  \\ \notag
  C_{kk} :=&\ \int^{1}_0 dy 2y \int^1_0 dx \frac{ (1-y)^2 }{\Delta_3},
\quad&               
  C_{pp} :=&\  \int^{1}_0 dy 2y \int^1_0 dx   \frac{y^2 (1-x)^2}{\Delta_3},
  \\ \notag
  C_{kp} :=&\ \int^{1}_0 dy   2y  \int^1_0 dx \frac{-y(1-y)(1-x)}{\Delta_3},  
 \quad& & 
\end{align}
with
\begin{align*}
   \Delta_3 :=  yx +  (1-y)r_a + y(1-x)r_b  - y(1-y)(1-x) r_q.     
\end{align*}
Here, the couplings are read as $y_X^\dag g_Y y_Z = y_X^{aj*} g_Y^{ab} y_Z^{bi}$
for $X,Y,Z = L,R$ and $a,b$ run over the internal fermions.

\subsection{Correction to photon coupling}

When the gauge boson is the photon, $g_L^{ab} = g_R^{ab} = e Q_e \delta^{ab}$ 
is independent
of the flavor of the internal fermion.  
In this case, the 1-loop amplitude after the wave function renormalization is given by 
\begin{align*}
  \Mcal^\mu 
  =: &
  \frac{eQ_e}{32\pi^2 m_S^2}  \ol{e}_j \sum_a \Biggl[
  (q^2\gamma^\mu - \dsq q^\mu ) H(r_a) \yy{L}{L}
  \\ \notag 
&\  -2 \left\{
  F(r_a) (m_j \yy{L}{L} + m_i \yy{R}{R}) 
  + m_a G(r_a)  \yy{R}{L}
  \right\} \sigma^{\mu\nu} q_\nu
  \Biggr]
 P_L e_i  + \left( L\leftrightarrow R \right), 
\end{align*}
where the functions of $r = m_a^2/m_S^2$ are given by
\begin{align*}
\label{eq-defHFG}
  H(r):=&\ - C_k -\frac{1}{2} C_{kk} = \frac{7r^3-36r^2+45r-16+6(3r-2)\log{r}}{18(1-r)^4},
  \\ \notag 
  F(r):=&\ -\frac{1}{2}C_k -\frac{3}{4} C_{kk} = \frac{2+3r-6r^2+r^3+6r \log{r}}{12(1-r)^4},
  \\ \notag
  G(r):= &\ - C_k = -\frac{3-4r+r^2+2\log{r}}{2(1-r)^3}. 
\end{align*}

For a lepton $\ell$, we match this result to the effective operator
~\cite{Dermisek:2022aec},
\begin{align*}
  \Lcal = \ol{\psi}_\ell \sigma^{\mu\nu} \Biggl(
     \frac{ie}{4m_\ell} \Delta a_\ell  
     + \frac{1}{2} d_\ell   \gamma_5  
          \Biggr)  \psi_\ell F_{\mu\nu}.
\end{align*}
We then obtain
\begin{align*}
  \Delta a_\ell =&\ -\frac{m_\ell^2 Q_\ell}{8\pi^2 m_S^2} \sum_a
  \Bigl[ (y_L^\dag y_L + y_R^\dag y_R ) F(r_a) 
  + \frac{m_a}{m_\ell} \mathrm{Re}\left( y_L^\dag y_R \right) G(r_a)\Bigr], 
  \\ \notag
  d_\ell =&\ -\frac{em_aQ_\ell}{16\pi^2 m_S^2} \sum_a \mathrm{Im}\left( y_R^\dag y_L \right) G(r_a),   
\end{align*}
where $Q_\ell = -1$ is the electric charge of the charged lepton.
We only study the magnetic dipole moment in the main text. 
The same amplitude also gives the decay width of the flavor-violating
process $e_i\to e_j\gamma$.

\subsection{Correction to $Z$ boson coupling}

We neglect terms of $\order{m_{i,j}}$
and the terms proportional to $p^\mu$ and $k^\mu$, which are irrelevant
for the decay rate.
The amplitude with wave function renormalization is
\begin{align}
  \delta \Mcal^\mu
  =&\ \frac{1}{32\pi^2} \ol{e}_j \gamma^\mu y_L^\dag
  \Biggl[
      \left(\frac{1}{\eta} - 1 - 2C_{00} - r_q C_{kp}  \right) g_R
      + \sqrt{r_ar_b} C_0 g_L
      - \left( \frac{1}{\eta} + B_1 \right) g_L^e
     \Biggr]
     y_L P_L e_i
 \notag      \\
   &\  + \left( L\leftrightarrow R \right).
\end{align}
In the LPDM model, the Yukawa and $Z$-boson couplings are given
by Eqs.~\eqref{eq-yLyR} and~\eqref{eq:Zc}. 
Hence, the divergent parts are canceled because 
\begin{align*}
&  y_L^\dag \left( g_R - g_L^e \right) y_L
  \propto \abs{\la_L}^2  \left[ H_L - 1  \right]_{11} = 0,
  \\
&  y_R^\dag \left( g_L - g_R^e \right) y_R
  \propto \abs{\la_E}^2  \left[ H_L - 0   \right]_{22} = 0.  
\end{align*}
The remaining finite part is given by
\begin{align}
  \label{eq:McalZ}
  \delta \Mcal^\mu
  =&\ \frac{1}{32\pi^2} \ol{e}_j \gamma^\mu \sum_{a,b} \Biggl[ 
       \la_L^{j*}\la_L^i \left( U_R \right)_{1a}
   \Biggl(
      H_{ab} g_R^{ab} + \ol{H}_{ab} g^{ab}_L
      + I_{a} g_L^e \delta_{ab}       
   \Biggr) 
     \left( U_R^\dag \right)_{b1} P_L
  \\ \notag
   & \hspace{2.0cm}+
      \la_E^{j}\la_E^{i*}
     \left( U_L \right)_{2a}
   \Biggl(
      H_{ab} g_L^{ab} + \ol{H}_{ab} g^{ab}_R
      + I_{a} g_R^e \delta_{ab}        
   \Biggr) 
     \left( U_L^\dag \right)_{b2} P_R \Biggr] e_i
  \\  
  =:&\ \frac{1}{32\pi^2} \ol{e}_j \gamma^\mu
     \Biggl\{ \left(\la_L^{j*}\la_L^i\right) F_L P_L
            + \left(\la_E^{j}\la^{i*}_E\right) F_R P_R \Biggr\} e_i,
        \label{eq-defFLFR}
\end{align}
with
\begin{align*}
  H_{ab} := -1-2C_{00}^{ab} - r_q C_{kp}^{ab},
  \quad 
  \ol{H}_{ab}:= \sqrt{r_ar_b} C_0^{ab},
  \quad
  I_{a} := -B_1^a(0).
\end{align*}
The functions $F_L$ and $F_R$ are defined in Eq.~\eqref{eq-defFLFR}.

{\small
\bibliography{ref} 
\bibliographystyle{JHEP} 
}

\end{document}